# "The Good, The Bad And The Ugly": Evaluation of Wi-Fi Steganography

Krzysztof Szczypiorski[1], Artur Janicki[1] and Steffen Wendzel[2]

[1] Warsaw University of Technology, Institute of Telecommunications, 15/19 Nowowiejska Street, 00-665 Warsaw, Poland

[2] Fraunhofer-Institut für Kommunikation, Informationsverarbeitung und Ergonomie FKIE, Fraunhoferstr. 20, 53343 Wachtberg, Germany

Email: ksz@tele.pw.edu.pl, a.janicki@tele.pw.edu.pl, steffen.wendzel@fkie.fraunhofer.de

*Abstract* —In this paper we propose a new method for the evaluation of network steganography algorithms based on the new concept of "the moving observer". We considered three levels of undetectability named: "good", "bad", and "ugly". To illustrate this method we chose Wi-Fi steganography as a solid family of information hiding protocols. We present the state of the art in this area covering well-known hiding techniques for 802.11 networks. "The moving observer" approach could help not only in the evaluation of steganographic algorithms, but also might be a starting point for a new detection system of network steganography. The concept of a new detection system, called MoveSteg, is explained in detail.

## I. INTRODUCTION

"Coma" is a classic suspense film based on the novel of the same name by Robin Cook. When a female surgery resident notices an abnormal amount of comas relative to the number of patients, she uncovers a horrible conspiracy: the place of her professional activity is not a real hospital, but rather (we are sorry for disclosing the end of this movie) a farm for healthy and fresh organ donors. The synopsis of this film shows that sometimes things are not always what they seem. This is sad and true, but this is also a paradigm of information hiding, when somebody uses a carrier to fetch a hidden message (as a steganogram) to cheat an observer, who is unable to guess that covert communication exists.

We may imagine an almost infinite number of carriers [1] depending on technology awareness and the power of the observers. In this paper we are focusing on the network as a carrier, so we are taking into account the network steganography [2], which are information hiding techniques that utilize network protocols as enablers of covert communication. We will divide a range of network steganography algorithms into three categories (inspired by the Italian epic spaghetti western film directed by Sergio Leone) depending on the undetectability by the observer: "good", "bad", and "ugly" algorithms. We assume a worst-case scenario for the sender (on the other hand it is the best for the observer): the observer is very close to the source of the steganograms. This assumption will allow recognition of the "good" and "bad" techniques. Later, we will consider when the observer has been moved away from the source, and some "bad" algorithms will be considered to be "ugly" ones. This is the basis of "the moving observer" technique that could help not only in the evaluation of steganographic algorithms, but also would be able to create fundamentals for a new detection system of network steganography called MoveSteg. The sender is observed from a distance by a group of sensors (observers), and some patterns of the traffic are evaluated from different perspectives to uncover any malicious behavior.

We chose Wi-Fi steganography as a solid, but still not fully explored, family of information hiding methods; the next section presents the state of the art in this research area. In Section III we explain the concept of "the moving observer" and evaluate the algorithms presented in the previous section. In Section IV we present the concept of MoveSteg – a new detection system of network steganography based on "the moving observer". Section V concludes this work and highlights future work.

## II. WI-FI STEGANOGRAPHY[1]: STATE OF THE ART

### A. Background

The family of IEEE 802.11 standards is the most popular method of wireless Internet access and the way to organize Small Office Home Office (SOHO) networks. Due to the popularity of Wi-Fi networks and their advantages, such as easy configuration and significant bandwidth, they also bring some risks. One of them is the simplicity of radio channel eavesdropping. Since Wi-Fi networks belong to the family of shared medium networks, their members have the possibility of "hearing" all the data frames exchanged in the air within the transmission range. From the point of view of network steganography, such a feature of broadcast communication (one to all) is very attractive, as it can allow numerous scenarios for hidden communication. It must be remembered, however, that for radio-based networks "the range is the limit". For Wi-Fi networks, typically the range is not larger than 100 meters (330 feet) from a source station, and it must be noted that techniques developed for Wi-Fi steganography can also be useful for hidden communication in "long-distance" wireless networks, like Long Term Evolution (LTE) or Worldwide Interoperability for Microwave Access (WiMax), which sounds very promising.

---

[1] This section is based on [3].

Below we review the most well-known information hiding techniques for wireless networks, while classifying them according to the characteristic features they exploit. We also highlight existing efforts to implement these steganographic methods. We divide this section as follows:
- Methods based on intentionally corrupted checksums.
- Methods using padding at physical layer.
- Methods based on controlling intervals between OFDM symbols.
- Other methods.
- Implementations of steganographic techniques.

### B. Methods Based on Intentionally Corrupted Checksums

The first steganographic system for Wi-Fi networks [4], according to our knowledge, was proposed by Szczypiorski and named HICCUPS (HIdden Communication system for CorrUPted networkS) [5]. For hidden transmission this method used frames with intentionally corrupted checksums. When a wireless client detected an error in a broadcast frame by checking the packet's checksum, the client simply dropped the corrupted frame. In HICCUPS these rejected frames were used for hidden transmission. To detect HICCUPS, one needs to observe the number of frames with incorrect checksums. If the number of those frames is statistically anomalous, then a hidden communication can be suspected. Another way of detecting HICCUPS is based on comparing the dropped and retransmitted frames. Szczypiorski [6] showed that HICCUPS bandwidth was as high as 1.27 Mb/s for an IEEE 802.11g 54 Mb/s network with 10 stations, assuming that 5% of traffic is used for steganographic transmission. Hardware implementation of HICCUPS is rather difficult because frame control sums are calculated on the board of the network cards. It is possible (but not yet proven) to build a HICCUPS-enabled card on top of a field-programmable gate array (FPGA) card with the capabilities of a software defined radio [7]. Najafizadeh *et al.* [8] presented a simulation of HICCUPS based on the initial work of Odor *et al.* in [9].

### C. Methods Using Padding at Physical Layer

The WiPad (Wireless Padding) [10] method was proposed by Szczypiorski and Mazurczyk for the 802.11a/g standards. It exploited the fact that some Wi-Fi networks (especially the newest ones) use orthogonal frequency-division multiplexing (OFDM). To send data over the radio channel this technique uses several narrow-bandwidth carriers of different frequencies. These carriers are more resilient to atmospheric conditions than a single wideband one, thus allowing for a more robust transmission. To minimize interference, OFDM divides the bits up into groups of pre-defined length, known as symbols, to assign them to the carriers. Most often the division of bits into symbols is not perfect: there will usually be some symbols left with too few bits. Therefore, there is so called "bit padding", i.e., OFDM transmitters adding extra throwaway bits, to make the symbols conform to the standard size. Since these bits are meaningless for the upper layers, they could be replaced with secret data. The analysis of the IEEE 802.11 frame showed that two other fields are also liable to padding: SERVICE and TAIL. The lengths of SERVICE and TAIL are constant (16 and 6 bits, respectively), while a Physical layer Service Data Unit (PSDU) is a medium access control (MAC) frame and its length depends on user data, ciphers, and network operation mode (ad hoc vs. infrastructure). The padding is present in all frames, therefore some frames that are more frequently exchanged, like ACKs (acknowledgments) may become an interesting target for covert communication. The OFDM padding study [10] led to similar trials with the LTE [11] and WiMax [12] technologies, researched by Grabska *et al.*

### D. Methods Based on Controlling Intervals Between OFDM Symbols

The work on WiPad was further continued and extended in [13] by Grabski and Szczypiorski. In this work the authors developed steganographic methods for high speed networks like 802.11n. They also exploited the fact that due to the multipath propagation present in radio transmission the receiver captures not only the signal propagated directly from the source, but also its delayed copies. As a consequence, the OFDM symbols can suffer from inter-symbol interferences. To mitigate this phenomenon a special safety interval can be inserted between OFDM symbols. Normally, in accordance with the IEEE 802.11 standards, these intervals are filled with the cyclic prefix. The system proposed by Grabski *et al.* was based on changing the prefixes of selected OFDM symbols and using them for hidden transmission. The steganographic bandwidth of such a covert channel ranged from 3.25 Mb/s to 19.5 Mb/s. A variation of this technique was implemented by Classen *et al.* [14].

### E. Other Methods

Other hiding techniques developed for Wi-Fi networks used CSMA/CA (Carrier Sense Multiple Access with Collision Avoidance) [15], controlling the rate of transmission [16], or a beacon mechanism [17].

Holloway [15] proposed a method called Covert DCF - a timing channel based on the CSMA/CA mechanism used in 802.11 to avoid collisions. The main idea is based on controlling the random backoff – an internal mechanism for counting the time until the next transmission. The author achieved a steganographic bandwidth of around 2 kb/s. Calhoun *et al*. [16] proposed a side channel that used the 802.11 MAC rate switching protocol to send hidden data between an access point and a station. The authors suggested two applications for this channel: covert authentication and covert Wi-Fi botnets. The covert authentication proposed by Calhoun *et al*. [16] was further developed by Sawicki and Piotrowski [17] to authenticate access points by using beacon frames and timestamp fields.

*F. Implementations of Steganographic Techniques*

Some of the steganographic channels were implemented in software [4][18][19]. Almost all of them were based on the modification of a header, which was the easiest way of controlling this part of the frame. Grabski and Szczypiorski [20] also proposed a tool for steganalysis. To our knowledge, only two publications have described hardware implementations, namely [7] and [14].

Kraetzer *et al*. [4] proposed two steganographic methods based on packet duplication/modification. The first method was based on header modification, especially on the "Retry" and "More Data" bits and on the "Duration/ID" field. The second technique exploited time dependencies between frames, and the authors revised their work in [21]. Frikha *et al*. [18] created a steganographic system based on header modification at the MAC sublayer. The authors proposed to use independently two fields of a MAC header: Sequence Control (SC) and Initial Vector (IV) for WEP (Wired Equivalent Privacy) protected frames. It is worth noting that WEP encryption is currently not commonly used as it is deemed insecure. Goncalves *et al*. [19] extended the work of Frikha *et al*. and proposed to use two bits in the protocol version field of the Frame Control Field in a MAC header to carry hidden information. Finally, Grabski *et al*. [20] implemented a Wi-Fi steganalyser as a tool to passively monitor the network traffic to detect hidden communication. The system was able to recognize five scenarios based on [4] and [18], including those based on the "Duration/ID" field as well as "Retry" and "More Data".

Dutta *et al*. [7] proposed a covert channel based on encoding covert information in the physical layer of common wireless communication protocols (called Dirty Constellation) and demonstrated the technique by implementing it in a software-defined radio for 802.11a/g.

Classen *et al*. [14] implemented four different covert channels:

- Using the Short Training Field in combination with Phase Shift Keying (PSK).
- Using the Carrier Frequency Offset with Frequency Shift Keying (FSK).
- Using 802.11a/g with additional subcarriers conforming to the 802.11n spectrum mask (an extension of [7]).
- Replacing parts of the OFDM Cyclic Prefix (an extension of [13]).

III. "THE MOVING OBSERVER" TECHNIQUE - EVALUATION OF THE ALGORITHMS

According to [22], network steganography communication can be characterized by four features: bandwidth, undetectability, robustness, and cost. The first three features, introduced by Fridrich in 1998 [23], are often presented as vertices of a triangle to show the interdependence among them. For example, the higher the required bandwidth, the more difficult it is to achieve high undetectability and robustness. This interdependence causes the need for a trade-off between the three features when a new steganographic system is designed. The last feature mentioned – cost – indicates the degradation of the carrier caused by the insertion of the secret data procedure [24]. It is important to note that the cost depends on each particular carrier, and it could be expressed in many ways, e.g. increased delay of packets, increased bit error rate, etc. It should be emphasized that the most desirable characteristic of network steganography communication is usually undetectability [25].

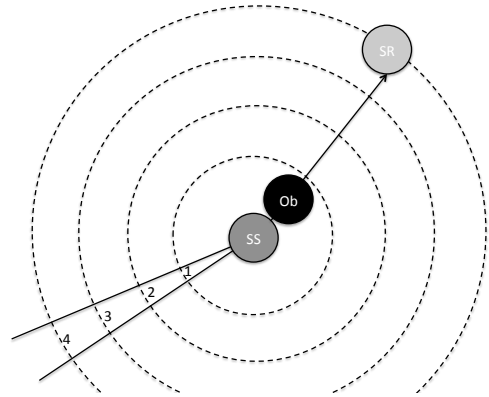

Fig. 1. The moving observer (Ob) is located close to the steganogram sender (SS).

"The moving observer" technique allows the evaluation of the strength of steganographic algorithms related to the undetectability. We considered three levels of undetectability named: "good", "bad", and "ugly" defined as follows:

- "Good" – the observer **is unable** to detect a hidden communication at the source of the steganograms (SS) – Fig. 1.
- "Bad" – the observer **is able** to detect a hidden communication at the source of the steganograms (SS), but he/she **is unable** to detect this communication, when he/she is moved away from the SS – Fig. 1.

- "Ugly" – the observer **is able** to detect a hidden communication anywhere in the network, even at the steganographic receiver (SR) – Fig. 2.

Always the proximity of the observer to the SS gives a chance to learn exactly the relations among the packets and their structure. Moving far away from the source involves phenomenon in the network like noise, which is introduced by the physical channel, but also by each switching hop adding some distortions on the inter-arrival time, size, or order of the packets.

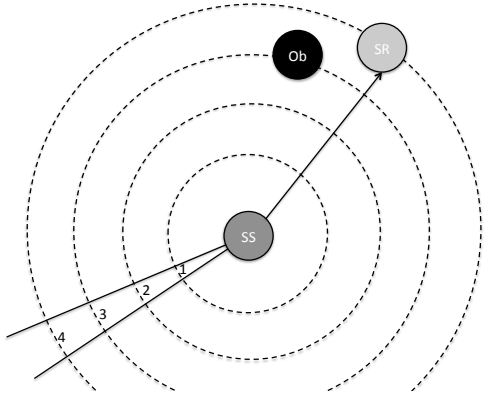

Fig. 2. The moving observer (Ob) is located close to the steganogram receiver (SR).

Fig. 1 and Fig. 2 present an abstract view of the network topology. This view is discrete rather than continuous – numbered circles (1-4) are similar to the orbits that could be equivalent to the hops (in logical order) or even geographical distances (in physical order, especially for Wi-Fi networks).

TABLE I: EVALUATION OF WI-FI STEGANOGRAPHY

| Author(s) and acronym (if exists) | Mark | | |
|---|---|---|---|
| | Good | Bad | Ugly |
| Calhoun et al. [1] | | | ✓ |
| Classen et al. [14] | | | ✓ |
| Dutta et al. [7] | | ✓ | |
| Frikha et al. [18] | | | ✓ |
| Goncalves et al. [19] | | | ✓ |
| Grabski et al. [13] | | | ✓ |
| Holloway [15] | ✓ | | |
| Kraetzer et al. [4] – first method | | | ✓ |
| Kraetzer et al. [4] – second method | ✓ | | |
| Sawicki et al. [17] | | | ✓ |
| Szczypiorski [5], HICCUPS | | ✓ | |
| Szczypiorski et al. [10], WiPad | | | ✓ |
| **Total marks** | **2** | **2** | **8** |

In a worst-case scenario for the sender (and the best for the observer): the observer is very close to the SS (Fig. 1), but typically he/she is somewhere in the network, so a best-case scenario for the sender (and worst for the observer) is to be very close to the SR (Fig. 2). We evaluated the proposals presented in the previous section according to the definitions of "good", "bad", and "ugly". Some of these proposals included several algorithms ([14] has four), but we distinguish specific algorithms only when it was necessary ([4] – first and a second methods). According to Table I, two proposals are marked as "good" ([15] and [4] – second method), two as "bad" ([5][7]), and eight as "ugly" ([1][10][13][14][17][18][19] and [4] – first method). "Ugly" methods are relatively naïve and could always be detected. "Bad" ones are only sophisticated for some distance from the source: both [5] and [7] could be recognized as a transmissions in a very noisy channel. The methods presented in [15] and [4] (second method) are "good" ones – both of these methods change the logical understanding of the Wi-Fi networks – [15] at the CSMA/CA level and [4] the interpretation of frame relations.

## IV. MOVESTEG: NEW DETECTION SYSTEM OF NETWORK STEGANOGRAPHY

"The moving observer" technique is a basis that can help not only in the evaluation of steganographic algorithms, but it also would be able to create the fundamentals for a novel network steganographic detection system called MoveSteg. The main target of the system is to detect "bad" methods. "Good" ones are strong enough to be detected even at the SS, so they are out of MoveSteg's scope. Detection of "ugly" methods can be made anywhere with sensors (observing nodes), as presented by Grabski et al. [20], and we assumed that the observer can break all these algorithms. Most interesting is cracking "the bad" methods, especially by moving from as close as possible to the SR to the SS or by observing SS from several perspectives (the minimum is two). Figures 3, 4, and 5 illustrate three scenarios:

- Two observing nodes ($N_1$ and $N_2$) are located close to SR, but none of these nodes are close to SS.
- Two observing nodes ($N_1$' and $N_2$') are located close to SR', one of them ($N_2$') is close to SS.
- An observing node $N_1$" is located close to SR".

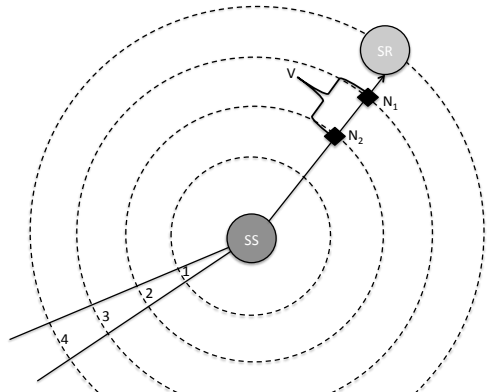

Fig. 3. Two observing nodes ($N_1$ and $N_2$) are located close to SR.

Fig. 3 shows that a hypothetical value V can be measured and compared from two nodes. For example, V could be a growth in delay – if it is almost 0, but a huge delay still exists from SS that could be an indicator of a delay based on steganography. In Fig. 4, V is called V'.

Fig. 6 presents the superposition of observing nodes and steganogram receivers that could be a way of cooperation to detect network steganography. Therefore, the values of V, V', and V'' are used together.

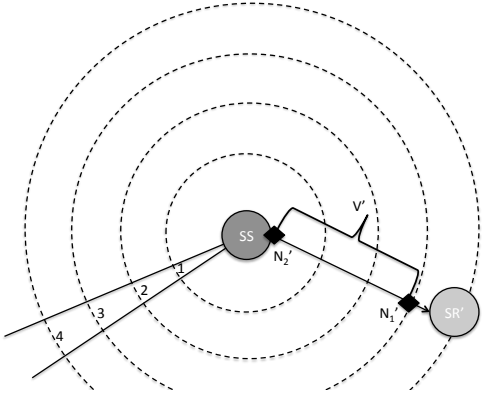

Fig. 4. Two observing nodes ($N_1'$ and $N_2'$) are located close to SR'; $N_2'$ is also close to SS.

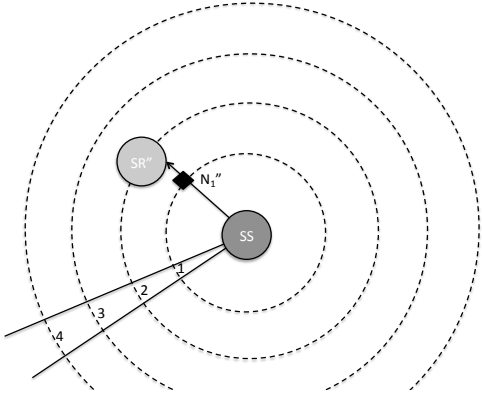

Fig. 5. An observing node $N_1''$ is located close to SR''.

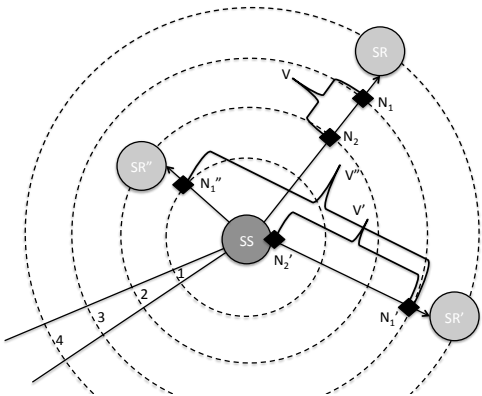

Fig. 6. Superposition of observing nodes and steganogram receivers.

V. CONCLUSIONS AND FUTURE WORK

The main aim of network steganography designers is to propose stealth methods linked to a minimized detectability. We introduced a categorization of such methods. "The good" methods are perfectly applicable and hardly detectable anywhere in a network, "the bad" ones are acceptable from some distance as they are detectable at the source of the steganograms, but "the ugly" ones are naïve and should be avoided.

In this paper we outlined the concept of MoveSteg applied to Wi-Fi networks, but it could be used with any type of the network. Future work will focus on the implementation of MoveSteg and is currently in progress.

ACKNOMLEDGMENTS

The concept of this paper was developed after fruitful discussion during Krzysztof Szczypiorski's seminar at George Mason University (GMU), Fairfax, Virginia (USA) in October 2014. Especially, Krzysztof Szczypiorski would like to thank Prof. Sushil Jajoida and Prof. Kris Gaj (both affiliated with GMU) for their valuable comments.

Krzysztof Szczypiorski was supported by the European Union in the framework of the European Social Fund through the Warsaw University of Technology Development Programme.

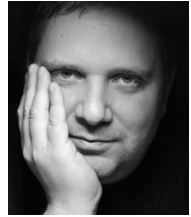
**Krzysztof Szczypiorski** is a Professor of Telecommunications at Warsaw University of Technology, Poland. He holds a DSc (habilitation, 2012), PhD (doctorate, 2007), and MSc (master's degree, 1997) all in telecommunications from WUT. He also finished his postgraduate studies in psychology of motivation (2013) at the University of Social Sciences and Humanities (SWPS), Warsaw, Poland. He graduated from Hass School of Business, University of California, Berkeley, USA (2013). Head and founder of Cybersecurity Division at the Institute of Telecommunications, WUT. A research leader of Network Security Group at WUT. His research interests include theory of observing change, network security, digital forensics, open-source intelligence, and wireless and ad-hoc communications. He is the author or the co-author of 150+ papers and 50+ invited talks. He is the inventor of two patents (one of them is pending). The guest editor of 10+ special issues of top ICT journals. For 20+ years he has served as an independent consultant in the fields of network security, telecommunications and computer science for many entities.

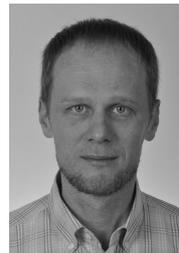
**Artur Janicki** received MSc and PhD degrees (1997 and 2004, respectively, both with honors) in telecommunications from the Faculty of Electronics and Information Technology, Warsaw University of Technology (WUT). Assistant Professor at the Institute of Telecommunications, WUT. His research and teaching activities focus on speech processing, including speaker recognition, speech coding and synthesis, and emotion recognition, with elements of data mining and information theory. Author of over 30 conference and journal papers, supervisor of over 40 bachelor and master theses. Member of the International Speech Communication Association (ISCA) and the European Association for Signal Processing (EURASIP).

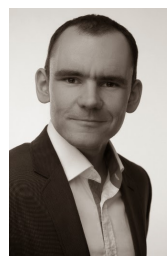
**Steffen Wendzel** studied computer science in Kempten and Augsburg, Germany. He received his PhD from the University of Hagen in 2013 and is the author of five books. After completion of his PhD he joined Fraunhofer FKIE where he is a head of a research team on smart building security.